\documentclass[prl,aps,superscriptaddress,footinbib,twocolumn]{revtex4-2}
\usepackage[utf8]{inputenc}
\usepackage[T1]{fontenc}
\usepackage{courier}
\usepackage{url}
\usepackage{xcolor}
\usepackage{amsmath}
\usepackage{amssymb}
\usepackage{stmaryrd}
\usepackage{graphicx}
\usepackage{bm}
\usepackage{dcolumn}
\usepackage{multirow}
\usepackage{esint}
\usepackage{ulem}
\usepackage{algorithm}
\usepackage{algorithmic}
\usepackage{float}
\usepackage{braket}
\usepackage{wasysym}

\DeclareMathOperator{\Tr}{Tr}
\usepackage[unicode=true,bookmarks=true,bookmarksnumbered=false,bookmarksopen=false,breaklinks=false,pdfborder={0 0 1},backref=false,colorlinks=true]
 {hyperref}
\hypersetup{
 linkcolor=magenta, urlcolor=blue, citecolor=red, pdfstartview={FitH}, hyperfootnotes=false, unicode=true}

\usepackage{todonotes}
\begin{document}
\title{
Demonstrating quantum error mitigation on logical qubits
}

% ====== affiliation ======
\newcommand{\zju}{School of Physics, ZJU-Hangzhou Global Scientific and Technological Innovation Center, \\and Zhejiang Key Laboratory of Micro-nano Quantum Chips and Quantum Control, Zhejiang University, Hangzhou, China}
\newcommand{\zjuoptical}{State Key Laboratory of Extreme Photonics and Instrumentation, \\College of Optical Science and Engineering, Zhejiang University, Hangzhou 310027, China}
\newcommand{\hefei}{Hefei National Laboratory, Hefei 230088, China}
\newcommand{\gscaep}{Graduate School of China Academy of Engineering Physics, Beijing 100193, China}

\author{Aosai Zhang} 
\thanks{These authors contributed equally.}
\affiliation{\zju}

\author{Haipeng Xie} 
\thanks{These authors contributed equally.} 
\affiliation{\gscaep}

\author{Yu Gao} 
\thanks{These authors contributed equally.}
\affiliation{\zju}

\author{Jia-Nan Yang} 
\affiliation{\zju}
\author{Zehang Bao}
\affiliation{\zju}
\author{Zitian Zhu}
\affiliation{\zju}
\author{Jiachen Chen}
\affiliation{\zju}
\author{Ning Wang} 
\affiliation{\zju}
\author{Chuanyu Zhang}
\affiliation{\zju}
\author{Jiarun Zhong}
\affiliation{\zju}

\author{Shibo Xu}
\affiliation{\zju}
\author{Ke Wang}
\affiliation{\zju}
\author{Yaozu Wu}
\affiliation{\zju}
\author{Feitong Jin}
\affiliation{\zju}
\author{Xuhao Zhu}
\affiliation{\zju}
\author{Yiren Zou}
\affiliation{\zju}
\author{Ziqi Tan}
\affiliation{\zju}
\author{Zhengyi Cui}
\affiliation{\zju}
\author{Fanhao Shen}
\affiliation{\zju}
\author{Tingting Li}
\affiliation{\zju}
\author{Yihang Han}
\affiliation{\zju}
\author{Yiyang He}
\affiliation{\zju}
\author{Gongyu Liu}
\affiliation{\zju}
\author{Jiayuan Shen}
\affiliation{\zju}
\author{Han Wang}
\affiliation{\zju}
\author{Yanzhe Wang}
\affiliation{\zju}

\author{Hang Dong}
\affiliation{\zju}
\author{Jinfeng Deng}
\affiliation{\zju}
\affiliation{\zjuoptical}
\author{Hekang Li}
\affiliation{\zju}
\author{Zhen Wang}
\affiliation{\zju}
\affiliation{\hefei}
\author{Chao Song}
\affiliation{\zju}
\affiliation{\hefei}
\author{Qiujiang Guo}
\affiliation{\zju}
\affiliation{\hefei}

\author{Pengfei Zhang}
\email{pfzhang@zju.edu.cn}
\affiliation{\zju}

\author{Ying Li}
\email{yli@gscaep.ac.cn}
\affiliation{\gscaep}

\author{H. Wang}
\affiliation{\zju}
\affiliation{\zjuoptical}
\affiliation{\hefei}

\begin{abstract}
A long-standing challenge in quantum computing is developing technologies to overcome the inevitable noise in qubits. 
To enable meaningful applications in the early stages of fault-tolerant quantum computing, devising methods to suppress post-correction logical failures is becoming increasingly crucial. 
In this work, we propose and experimentally demonstrate the application of zero-noise extrapolation, a practical quantum error mitigation technique, to error correction circuits on state-of-the-art superconducting processors. 
By amplifying the noise on physical qubits, the circuits yield outcomes that exhibit a predictable dependence on noise strength, following a polynomial function determined by the code distance. 
This property enables the effective application of polynomial extrapolation to mitigate logical errors. Our experiments demonstrate a universal reduction in logical errors across various quantum circuits, including fault-tolerant circuits of repetition and surface codes. 
We observe a favorable performance in multi-round error correction circuits, indicating that this method remains effective when the circuit depth increases. 
These results advance the frontier of quantum error suppression technologies, opening a practical way to achieve reliable quantum computing in the early fault-tolerant era. 
\end{abstract}

\maketitle

\section{Introduction}

Suppressing errors is a problem that lies at the center of quantum computing technologies. Quantum error correction and mitigation are the two generic methods of error suppression. Error correction promises to reach an arbitrarily high fidelity provided sufficient qubit resources. Recently, experiments have demonstrated a positive gain in the surface code error correction when scaling the code distance up to 7~\cite{AcharyaZobrist2024}. However, it is still viewed as a long-term goal to achieve negligible infidelity through error correction, which could need millions or even more qubits and pose a challenge to the experimental technologies regarding scalability~\cite{Preskill2018}. 
%The eventual solution to the scalability issue may even require a non-trivial update on the platform from current single-chip quantum processors to a distributed multi-chip quantum processor. 
Error mitigation takes different resources, computing time, to suppress errors, being originally designed for the regime that error correction is lacking~\cite{RevModPhys.95.045005}. Considering the earliest applications of quantum computing, it is highly likely that they will be carried out under strict technology constraints, mainly on the qubit number. In this scenario, error correction can only attain a limited fidelity, and a practical approach taking advantage of the two error suppression methods is necessary~\cite{PiveteauTemme2021,SuzukiTokunaga2022}. 

Zero-noise extrapolation (ZNE) is one of the most practical error mitigation techniques, universally applicable to quantum algorithms that evaluate expectation values~\cite{PhysRevX.7.021050,PhysRevLett.119.180509}. The central idea behind ZNE is to amplify the noise in a quantum circuit by a controllable factor $r$, and then extrapolate the results back to $r = 0$, thereby inferring the behavior of a noiseless circuit. In a recent experiment, ZNE demonstrated substantial error suppression on a superconducting system with more than a hundred qubits~\cite{KimKandala2023}, making it the only error mitigation technique successfully applied at this scale to date. Given these promising results, we consider ZNE the leading candidate for mitigating errors in quantum error correction circuits. Specifically, we apply ZNE to reduce post-correction logical errors by amplifying noise on physical qubits. 

The implementation of ZNE on logical qubits faces two primary experimental challenges. First, it requires a quantum processor with high-fidelity gates and sufficient qubits, which is crucial for executing quantum error correction. Recent progress in qubit fabrication and control technologies have made superconducting qubits a promising platform for investigating quantum error correction and mitigation technologies~\cite{KimKandala2023,OBrienRubin2023,vandenBergTemme2023,XuDeng2023,KrinnerWallraff2022,ZhaoPan2022,AcharyaZobrist2024}. In this work, we utilize two superconducting quantum processors to experimentally assess residual errors and costs in quantum error correction and mitigation, employing both the repetition code and surface code~\cite{DennisPreskill2002,FowlerSurface2012,DevittNemoto2013,Terhal2015}. Second, accurate error mitigation relies on measuring error rates per operation~\cite{RevModPhys.95.045005,van_den_berg_probabilistic_2023}. While measuring logical error rates is feasible for small codes, it becomes increasingly time-consuming for larger codes due to the small logical error rates~\cite{eisert_quantum_2020,PhysRevA.105.032435}. Additionally, for high-encoding-rate quantum error correction codes, such as certain qLDPC codes, logical errors may involve multi-qubit correlations, making the measurement impractical~\cite{PRXQuantum.2.040101,bravyi_high-threshold_2024} (see Supplementary Section~S2). To overcome this issue, we adopt a strategy of amplifying noise on physical qubits instead of logical qubits. By integrating error mitigation techniques with error correction, this study demonstrates a practical pathway to bridge the gap between the noisy intermediate-scale quantum (NISQ) era and the fault-tolerant quantum computing (FTQC) era, advancing the pursuit of practical quantum computing technologies.

\section{Results}

We shall first justify the application of ZNE to quantum error correction circuits. The essential assumption made in ZNE is that the expected value of an observable is a function $\langle O\rangle(r)$ of the noise strength $r$, and this function can be fitted well by simple functions such as polynomials. This assumption holds in NISQ circuits, which has been illustrated in many theoretical works and experiments~\cite{PhysRevX.7.021050,PhysRevLett.119.180509,RevModPhys.95.045005,Czarnik2021errormitigation,PRXQuantum.2.040330,KandalaGambetta2019,KrinnerWallraff2022}. In this paper, one of the primary goals is to demonstrate that the assumption is also valid in quantum error correction circuits. 

\begin{figure}[tbp]
\centering
\includegraphics[width=3.5 in]{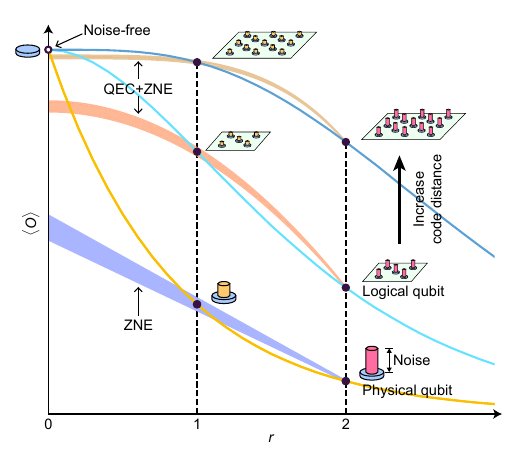}
\caption{
{\bf Schematic diagram of ZNE on logical qubits.} For an arbitrary quantum circuit with or without error correction, we can amplify the noise in either all physical operations or those causing most errors to mitigate corresponding imperfections. When using the circuit to measure the expectation of an observable $O$, the value changes with the noise strength $r$, depicted by a function $\langle O\rangle(r)$ for each case (lines). To implement a $K$th-order ZNE, we amplify the noise in the circuit and measure the observable expectation $\langle O\rangle$ at $K+1$ different noise strengths $r_0,r_1,\ldots,r_K$ ($K = 1$ in the figure). We always take $r_0 = 1$ corresponding to the raw noise without amplification. With the expectation values $\langle O\rangle(r_0),\langle O\rangle(r_1),\ldots,\langle O\rangle(r_K)$ (black solid circles), we fit the function and infer the noise-free result $\langle O\rangle_{\rm ideal}$ using a $(K+1)$-term polynomial (shadows, where the width indicates the variance in inference). We choose different polynomials depending on whether error correction is utilized and the code distance. The combination of error correction and ZNE results in a smaller bias and variance of observable expectations. 
}
\label{fig:illustration}
\end{figure}

Our first evidence is a theoretical result based on the stochastic error model: each operation in the circuit is either error-free or erroneous with a certain probability of $p$, and we increase the probability to $rp$ (with $r>1$) when amplifying the noise (Fig.~\ref{fig:illustration}). We apply this model to general quantum circuits with all the components necessary for quantum error correction, including mid-circuit state preparation and measurement, feedback and post-selection; these components are beyond NISQ circuits~\cite{chen_complexity_2023}. Even with these components, the expected value with errors is in the form $\langle O\rangle(r) = \langle O\rangle_{\rm ideal} + \sum_{k=1}^N a_kr^k$, where $N$ is the number of operations, and $a_k$ is the deviation caused by $k$ errors in the circuit (see Supplementary Section~S1). This polynomial expression holds for all quantum circuits and suggests the use of a polynomial fitting function. 

We use different polynomial fitting functions in cases with and without error correction. Without error correction, the leading contribution of errors is $a_1$ due to only one error in the circuit. Therefore, we always include the linear term $a_1r$ when choosing a polynomial fitting function, then we introduce $r^2, r^3, \ldots$ terms sequentially for higher-order fitting; we use such polynomials in conventional ZNE. With error correction, the leading order is different. In a circuit, its capability of error correction is characterized by an effective code distance $d$: when the number of errors is smaller than $\lceil d/2\rceil$, we can always successfully detect and correct the errors. Because of this reason, the coefficients $a_k$ are all zero for $k = 1, 2, \ldots, \lceil d/2\rceil-1$, and the leading contribution becomes $a_{\lceil d/2\rceil}$; this coincides with numerical results on logical error rates of surface codes and bivariate bicycle codes~\cite{BravyiSimulation2013,bravyi_high-threshold_2024}. Therefore, with error correction, we choose a fitting function in the form $\langle O\rangle(r) = \langle O\rangle_{\rm em} + \sum_{k=\lceil d/2\rceil}^{\lceil d/2\rceil+K-1} a_k' r^k$, which has $K$ non-constant terms starting with the $r^{\lceil d/2\rceil}$ term.  

\begin{figure*}[tbp]
\centering
\includegraphics[width=\textwidth]{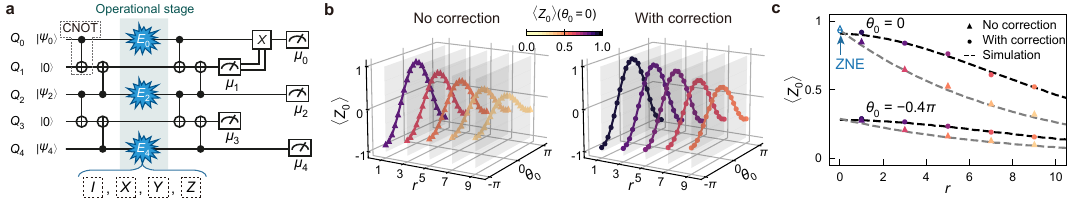}
\caption{
{\bf ZNE on an example circuit with the feedback error correction.}
{\bf a}, Circuit schematic. Each data qubit $Q_j$ is initialized into a superposition state parametrized by the rotation angle $\theta_j$. Either idling gate or one of the Pauli gates is injected as errors during the operational stage, effectively yielding 4 types of circuit instances. All qubits are measured simultaneously, returning bit strings for uncorrected data. Based on the bit strings of all circuit instances, post-selection on $Q_3$ and feedback $X$ gate are done numerically for corrected data. 
{\bf b}, Uncorrected (left) and corrected (right) expectation values of $Z_0$ observable as a function of rotation angle $\theta_0$ and noise scaling factor $r$, where the unit error probability is chosen as $p=8.8\%$.
Lines and markers are colored according to the measured values at $\theta_0 = 0$.
{\bf c}, ZNE for $\theta_0=0$ and $-0.4\pi$.
Dashed lines are numerical simulation results using a model with gate and measurement errors.
Empty symbols indicate the extrapolated results of ZNE using two data points at $r=1$ and $r=3$.
}
\label{fig:state_preservation}
\end{figure*}

We carry out experimental demonstrations on two superconducting quantum processors to certify that ZNE can be naturally integrated into the FTQC circuits. The processors used here, each of which contains a lattice of tens of frequency-tunable qubits featuring adjustable nearest-neighbor couplings, have performance similar to those dedicated to quantum error correction experiments in the literature~\cite{ZhaoPan2022,KrinnerWallraff2022,AcharyaZobrist2024}. The qubits selected for the experiments are highly coherent, with the median $T_1$ time being above $100$~$\mu$s (see parameters in Supplementary Table~S1), and the controlled $\pi$-phase (CZ) gates between nearest-neighbor qubits have a median fidelity around 0.995. 
Since FTQC also relies heavily on measurement quality, we can differentiate between the $\ket{0}$ and $\ket{1}$ states of each qubit, achieving median readout fidelities of $0.995$ and $0.991$ for Processor~\MakeUppercase{\romannumeral 1} (with Purcell filters, $0.5$-$\mu$s measurement time) and Processor~\MakeUppercase{\romannumeral 2} (without Purcell filters, $2.5$-$\mu$s measurement time), respectively.
This is accomplished by pumping the $\ket{1}$ state of each qubit to its next-higher level. 
Furthermore, even without the pumping technique which is a prerequisite for feedback operations in FTQC circuits, we can still achieve a median readout fidelity of $0.991$ on Processor~\MakeUppercase{\romannumeral 1} within 0.5~$\mu$s, as demonstrated in our repetition code experiment by allowing repetitive measurements on the syndrome qubits up to $M=4$ rounds (see next).

As the first experimental demonstration, we show that ZNE works on an example circuit with a feedback $X$ control to eliminate the bit-flip error on $Q_0$ (Fig.~\ref{fig:state_preservation}{\bf a}), where the nominal data qubits ($Q_j$ for $j=0$, 2, and 4) are each initialized into a superposition state given by $\ket{\psi_j} = \cos\dfrac{\theta_j}{2}\ket{0} - i\sin\dfrac{\theta_j}{2}\ket{1}$.
The first 4 CNOT gates in the sequence diagram are used to encode the parity of the data qubits onto the syndrome qubits ($Q_1$ and $Q_3$), followed by operations for algorithmic purpose, and the next 4 CNOT gates serve to decode and identify the bit-flip type of errors that may occur during the operational stage. To implement ZNE, one needs to be able to controllably amplify the errors, which can be achieved using schemes such as pulse stretching~\cite{TemmeGambetta2017,KandalaGambetta2019} or subcircuit repetition~\cite{HeBauer2020}.
Here to simulate depolarizing errors occurring at a probability of $rp$ with $r$ being the scaling factor, we run $4^3$ types of circuit instances which correspond to all combinations of inserting one operation drawn from the list of $\left\{I, X, Y, Z \right\}$ at the operational stage for each of the 3 data qubit, based on which we construct a weighted list of circuit instances so that the insertion probabilities of $X$, $Y$, and $Z$ all equal to $rp/3$ for each data qubit (see Supplementary Section~S4 for more details). 
Finally all qubits are measured in the $Z$ basis to yield a 5-bit binary string, $\mu_0\dots\mu_4$, for each circuit instance, and the no correction data are directly calculated using $\mu_0$s from all circuit instances.
In the case of error correction, since a bit-flip error due to the insertion of $X$ or $Y$ on the pair of data qubits $Q_j$ and $Q_{j+2}$, $j\in\{0,2\}$, flips the state of $Q_{j+1}$, 
the 5-bit binary strings from all instances are selected only when $\mu_3=1$ to ensure no error on $Q_2$, following which $\mu_0$ is numerically reversed due to the feedback $X$ gate only if $\mu_1=-1$ to eliminate the bit-flip error on $Q_0$. The newly processed strings are then used to calculate the data with correction.

For the case of no correction, Fig.~\ref{fig:state_preservation}{\bf b} shows that the expectation values of $Z_{0}$ for $Q_0$, for different input states parametrized by $\theta_0$, all gradually approach zero as the noise strength $r$ increases. In comparison, the data using processed bit strings with error correction indicate that the impact of noise is reduced, resulting in a slower decline in $\langle Z_0 \rangle$. Experimental results closely match numerical simulations of circuits with gate and measurement errors (Fig.~\ref{fig:state_preservation}{\bf c}). Furthermore, when we use ZNE to extrapolate the values at $r = 0$, the results show excellent agreement with numerical predictions, which demonstrates that ZNE effectively works on the circuit. 

In the above-mentioned example circuit, small errors occurring at the CNOT gates and measurement cannot be corrected, so that the post-mitigation residual error at $r=0$ is comparable to the case without error correction. 
Next, we resort to fault-tolerant circuits of repetition and surface codes, where the logical error rates can be suppressed to an arbitrarily low level by increasing the code distance once physical error rates are below the threshold. On fault-tolerant circuits, we find that the post-mitigation residual error is much smaller than the case without error correction. 

\begin{figure*}[tbp]
\centering
\includegraphics[width=\textwidth]{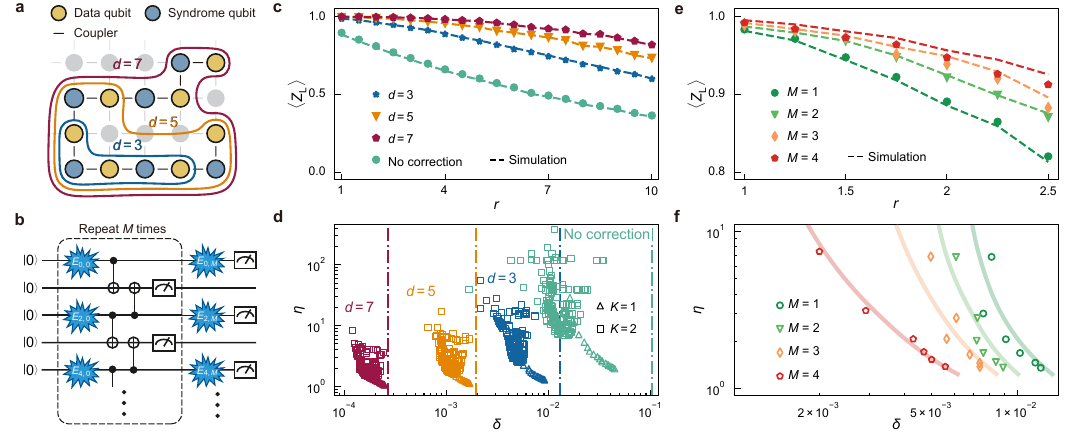}
\caption{
{\bf Performance of ZNE in the repetition code.}
{\bf a}, Layout of the repetition code on Processor~\MakeUppercase{\romannumeral 1}, with the data (golden) and syndrome (blue) qubits depicted as circles.
{\bf b}, Schematic of the experimental circuit, where error injection and parity measurement are repeatedly performed for $M$ rounds, followed by the final round of error injection and data qubit measurement.
{\bf c}, Measured expectation values of $Z_{\rm L}$ with respect to the noise scaling factor $r$ in circuits with one round of mid-circuit parity measurements. 
We present the uncorrected results (averaged over all data qubits) exclusively for the distance-7 repetition code, as the results for different code distances are nearly identical (Supplementary Fig.~S5). 
Dashed lines: numerical simulation results.
{\bf d}, Scatter plot showing the bias $\delta$ and sampling overhead $\eta$ for all possible choices of $r_1,\ldots,r_K$ in ZNE. 
Generally, the ZNE results of $K=2$ (square) exhibit reduced bias yet introduce higher sampling overhead compared to $K=1$ (triangle).
Dashed lines: biases without ZNE, i.e., $\delta_0 = \left| \langle Z_{\rm L} \rangle_{\rm ideal} - \langle Z_{\rm L} \rangle \left(r=1\right) \right|$.
{\bf e}--{\bf f}, Similar to panels {\bf c} and {\bf d}, but utilizing a multi-round repetition code with $d=7$ and performing ZNE with $K=1$. 
The lines in panel~{\bf f} serve as guides to illustrate the respective trends of ZNE performance.
}
\label{fig:repetition_code}
\end{figure*}

We implement repetition codes with distances $d=3, 5, 7$ using up to 13 qubits on Processor~\MakeUppercase{\romannumeral 1}, as shown in Fig.~\ref{fig:repetition_code}{\bf a}. Repetition codes protect the encoded logical qubit from bit-flip errors and have been widely used as a prototype for demonstrating error correction across various quantum platforms, including ion traps~\cite{ChiaveriniWineland2004,MosesPino2023}, nuclear magnetic resonance~\cite{CorySomaroo1998,MoussaLaflamme2011}, and superconducting qubits~\cite{ReedSchoelkopf2012,AcharyaZobrist2024}. Here, we adopt it for our first demonstration of error mitigation on fault-tolerant circuits. 

The circuit of a repetition code, schematically depicted in Fig.~\ref{fig:repetition_code}{\bf b}, consists of $d$ data qubits and $d-1$ syndrome qubits arranged alternately in a linear chain. The data qubits are initialized in the state $\ket{0}$, preparing the logical qubit in the logical state $\ket{0_{\rm L}}$. Next, we apply error injections and perform parity-check measurements for $M$ successive rounds. The circuit terminates with an additional round of error injection and transversal measurement on data qubits. The transversal measurement facilitates a final round of parity checks and readout of the logical operator $Z_{\rm L}$. The parity checks are designed to detect all possible bit-flip errors, whether injected deliberately or introduced by computational operations such as gates, state preparation, and measurements. Each experimental shot produces a bit string of length $M(d-1)+d$, recording the measurement outcomes. This bit string is then processed by a minimum-weight perfect matching (MWPM) decoder~\cite{Fowler2015,Gidney2021}, which outputs the correction gates. These corrections ensure faithful reconstruction of the logical state, provided that the number of bit-flip errors does not exceed $\lceil d/2\rceil-1$; therefore, the circuit is fault-tolerant with respect to bit-flip errors. 

We evaluate the performance of error suppression methods using the expectation value of the logical operator $Z_{\rm L}$. In an ideal noise-free circuit, the expectation value is $\langle Z_{\rm L} \rangle_{\rm ideal} = 1$; the presence of noise reduces this value. For illustrative purpose, we experimentally inject Pauli errors during the operational stage with the unit probability set at $p=3.6\%$, as done in the previous example. Figure~\ref{fig:repetition_code}{\bf c} shows the experimental results for circuits with one round of parity-check measurements ($M=1$): the expectation value decreases gradually with increasing noise strength, while error correction slows the rate of decline. As the code distance $d$ increases, the rate of decline approaches zero asymptotically, indicating a below-threshold error regime. 

To further suppress errors beyond error correction, we apply ZNE to the results. Specifically, we select $K+1$ data points corresponding to noise strengths $r=r_0, r_1, \cdots, r_{K}$ and perform polynomial extrapolation to obtain the error-mitigated expectation value of the logical operator, denoted as $\langle Z_{\rm L} \rangle_{\rm em}$. To evaluate the performance of ZNE, we define two metrics: the bias $\delta$ and the sampling overhead $\eta$. The bias 
$\delta$ quantifies the deviation of the error-mitigated value from the ideal expectation and is expressed as 
\begin{equation}
\delta = \left|\langle Z_{\rm L} \rangle_{\rm em} - \langle Z_{\rm L} \rangle_{\rm ideal} \right|.
\end{equation}
The sampling overhead $\eta$ measures the relative increase in sampling cost required to achieve the same variance for the error-mitigated value as for the uncorrected value, and it is defined as 
\begin{equation}
\eta = \dfrac{{\rm Var}\left[\langle Z_{\rm L} \rangle_{\rm em} \right]}{{\rm Var}\left[\langle Z_{\rm L} \rangle\left(r_0\right)\right]}.
\end{equation}
Unless otherwise specified, we take $r_0 = 1$ in what follows and iterate over all other data points for $r_1, \cdots, r_{K}$. 

The $\delta$-$\eta$ scatter plots in Fig.~\ref{fig:repetition_code}{\bf d} visualize that ZNE consistently provides more accurate estimations compared to the uncorrected value at $r_0 = 1$ (indicated by dashed lines) across all cases.
The plots also highlight a trade-off between the bias $\delta$ and sampling overhead $\eta$: while ZNE reduces bias, it increases the sampling overhead. If the noise-boosted data points at $r_1, \cdots, r_{K}$ are chosen closer to $r_0$ or if a higher-order extrapolation (larger $K$) is used, the bias generally decreases, but the sampling overhead tends to grow. 
Figures~\ref{fig:repetition_code}{\bf c-d} further highlight the advantage of combining ZNE with error correction. When error correction is applied, the $\delta$-$\eta$ scatter points progressively migrate toward the lower-left corner as the code distance $d$ increases, indicating simultaneous improvements in precision and efficiency. Notably, for $d=7$, the residual error $\delta$ is reduced to approximately $1 \times 10^{-4}$, with a modest sampling overhead of just $5$. 

It is important to note that error mitigation is scalable on error correction circuits as the code distance and circuit complexity increase~\cite{SuzukiTokunaga2022}. A fundamental limitation of error mitigation methods is that they become inefficient when the product of the error rate per gate and the number of gates becomes large. This limitation is well-studied in probabilistic error cancellation, a bias-free method, where variance increases exponentially with the gate number~\cite{takagi_fundamental_2022,quek_exponentially_2024,PhysRevLett.131.210601}. In contrast, the polynomial-function ZNE has a bias that grows with the gate number~\cite{PhysRevLett.119.180509}. Although these issues also arise in error mitigation applied to error correction circuits, the key factors become the logical error rate and the number of logical gates. Therefore, as long as the logical error rate remains sufficiently low --- that is, with a sufficiently large code distance --- error mitigation can be applied to circuits of arbitrary gate complexity. For instance, using the surface code with a physical error rate of $1\permil$ per gate and a code distance of eleven, we can achieve a logical error rate of approximately $2\times 10^{-10}$~\cite{BravyiSimulation2013}. This allows a circuit with $5\times 10^7$ logical operations to run at one logical error in one hundred circuit shots. In this case, our protocol can reduce the logical error by a factor of about $0.025$ with an overhead cost of $\eta \simeq 136$ due to our estimation (see Supplementary Section~S7). Moreover, our experiments show that as the code distance and parity-check rounds increase, the performance of ZNE remains nearly unchanged; notice that multi-round parity checks are necessary in practical fault-tolerant quantum computing. This insight can be demonstrated in two complementary ways. First, when the error rate per parity-check round is fixed, increasing the number of parity-check rounds results in a similar relative bias and sampling overhead for ZNE, despite the growth in unmitigated bias (see Supplementary Fig.~S6). Second, for a fixed total error rate in all rounds, partitioning the circuit and employing multi-round QEC significantly improves both the unmitigated and mitigated results (Fig.~\ref{fig:repetition_code}{\bf e}--{\bf f}). Remarkably, these enhancements are achieved while the sampling overhead remains nearly unchanged. 

\begin{figure*}[!htbp]
\centering
\includegraphics[width=\textwidth]{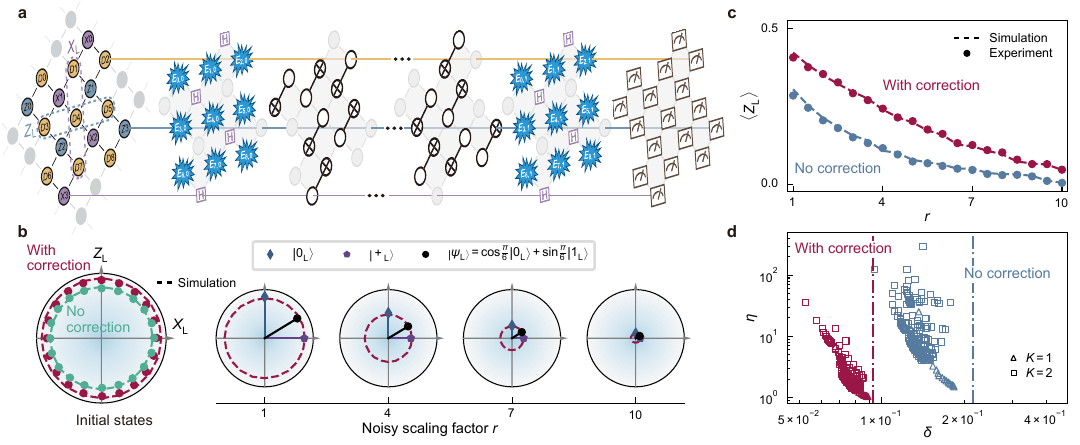}
\caption{
{\bf Correction and mitigation of both $X$- and $Z$-type errors in a distance-3 surface code.}
{\bf a}, Illustration of the experimental circuit implemented on Processor~\MakeUppercase{\romannumeral 2}. 
The qubit layout (left) comprises 9 data qubits (yellow), 4 $Z$-type syndrome qubits (blue), and 4 $X$-type syndrome qubits (violet).
The circuit consists of two error injection layers, interleaved with multiple layers of Hadamard gates (H) and CNOT gates (black) designed to perform parity check, and terminates with simultaneous readout on all data and syndrome qubits. The system is initialized in an arbitrary logical state of the surface code, with detailed protocol provided in Supplementary Section~S5.
{\bf b}, Logical states on the $X_{\rm L}$-$Z_{\rm L}$ plane of the Bloch sphere, constructed from the measured expectation values of $X_{\rm L}$ and $Z_{\rm L}$. The first plot shows various logical states measured without error injection, while the subsequent plots show the effect of different noisy scaling factors $r$ for three initial states: $|0_{\rm L}\rangle$, $|+_{\rm L}\rangle$ and $|\psi_{\rm L}\rangle=\cos\frac{\pi}{6}|0_{\rm L}\rangle + \sin\frac{\pi}{6}|1_{\rm L}\rangle$. 
{\bf c}, Expectation values of the $Z_\mathrm{L}$ observable for the logical state $|\psi_\mathrm{L}\rangle$ as a function of noise scaling factor $r$, including the raw data (blue points), corrected data (crimson points), and numerical simulation results (dashed lines).
{\bf d}, Scatter plot of the bias $\delta$ and sampling overhead $\eta$. The bias $\delta$ derived from the corrected data approaches the limit set by the imperfect initial state, which is close to $5\times 10^{-2}$.
}
\label{fig:surface_code}
\end{figure*}

As a final experiment, we extend our investigation to the rotated surface code, a quantum error correction code capable of correcting both bit-flip and phase-flip errors. 
We implement the distance-3 rotated surface code on our Processor~\MakeUppercase{\romannumeral 2}, as illustrated in Fig.~\ref{fig:surface_code}{\bf a}. The encoded logical qubit is defined by two anti-commuting logical operators, $X_{\rm L}=X_{D1} X_{D4} X_{D7}$ and $Z_{\rm L}=Z_{D3} Z_{D4} Z_{D5}$, and the system can be initialized by preparing the logical states using digital quantum circuits that consist of single- and two-qubit gates~\cite{SatzingerRoushan2021,ZhaoPan2022}. 

We demonstrate the effectiveness of ZNE following the sequence diagram shown in Fig.~\ref{fig:surface_code}{\bf a}, with three representative initial states: $ \ket{0_{\rm L}} $, $ \ket{+_{\rm L}} $, and $ \ket{\psi_{\rm L}} = \cos\frac{\pi}{6}\ket{0_{\rm L}} + \sin\frac{\pi}{6}\ket{1_{\rm L}} $ (Fig.~\ref{fig:surface_code}{\bf b}).
In the absence of error injections, the initial states reconstructed from the corrected expectation values of the $ Z_{\rm L} $ and $ X_{\rm L} $ observables locate significantly closer to the surface of the Bloch sphere compared to those reconstructed from the uncorrected values (Fig.~\ref{fig:surface_code}{\bf b}).
This demonstrates reductions in both bit-flip and phase-flip errors that arise during the initial state preparation and parity measurement processes.
We present numerical simulations based on the single-qubit depolarizing model, where the depolarizing rate is calibrated by matching the fidelity of the prepared logical state, $\ket{0_{\rm L}}$.
As shown in Figs.~\ref{fig:surface_code}{\bf b-c}, this noise model agrees well with the experimental data, regardless of the initial states.
Detailed results for both uncorrected and corrected expectation values of the $Z_{\rm L}$ observable, with the initial state set to $\ket{\psi_{\rm L}}$, are presented in Fig.~\ref{fig:surface_code}{\bf c}; see Supplementary Fig.~S8 for more experimental results.
To implement ZNE, we insert random Pauli gates on data qubits before and after the parity-check measurements to amplify the noise. In surface codes, quantum computing are driven by the parity-check measurements applied on a lattice deforming with time in protocols such as braiding transformation~\cite{FowlerSurface2012} and lattice surgery~\cite{Horsman_2012}. ZNE can be applied to such circuits through the same noise amplification strategy as taken in our experiment. 
By incorporating ZNE, the residual errors in $\langle Z_{\rm L} \rangle$ and $\langle X_{\rm L} \rangle$ are further reduced beyond error correction (see Fig.~\ref{fig:surface_code}{\bf d} and Supplementary Fig.~S8).

\section{Discussion}

Our results demonstrate for the first time the application of quantum error mitigation on error correction circuits, effectively minimizing the impact of error correction failures. The specific method that we choose is ZNE: we amplify noise on physical qubits and utilize a selected polynomial extrapolation function, thereby adapting the method to be compatible with fault-tolerant quantum circuits. Amplifying physical error rates, a well-established and experimentally validated technique on NISQ circuits, ensures the feasibility of our approach. Moreover, the success of error mitigation methods such as ZNE and probabilistic error cancellation fundamentally depends on precise noise modeling. Our experimental results exhibit strong consistency with numerical simulations of noise models from calibration, offering a robust foundation for further performance optimization. As the main result, our approach achieves a reduction in error-induced bias, outperforming both standalone error correction and error mitigation. This bias reduction illustrates the potential to achieve reliable quantum computing on hardware that only permits limited error correction capabilities, marking a pivotal step toward large-scale quantum computing on noisy devices. 

\vskip 0.5cm
\noindent{\bf Acknowledgements}

The device was fabricated at the Micro-Nano Fabrication Center of Zhejiang University. We acknowledge the support from the National Natural Science Foundation of China (Grant Nos.~92365301, 92065204, 12404574, 12274368, 12274367, 12174342, 12322414, 12404570, U20A2076, 12225507 and 12088101), the Innovation Program for Quantum
Science and Technology (Grant No.~2021ZD0300200), the Zhejiang Provincial Natural Science Foundation of China (Grant Nos.~LR24A040002 and LDQ23A040001), the National Key Research and Development Program of China (Grant No.~2023YFB4502600), and the NSAF (Grant No.~U1930403). 

\vskip 0.5cm
\noindent {\bf Author contributions}

H.X. and Y.L. proposed the ideas and conducted the theoretical analysis; A.Z., Y.G. and J.Y. carried out the experiments and analyzed the experimental data under the supervision of P.Z. and H.W.; H.L. and J.C. fabricated the device, supervised by H.W.; All authors contributed to the experimental setup, analysis of data, discussions of the results and writing of the manuscript.

\bibliography{references}

\clearpage

\setcounter{table}{0}
\renewcommand{\thetable}{S\arabic{table}}%
\renewcommand{\thefigure}{S$\the\numexpr\value{figure}-4$}%
\setcounter{equation}{0}
\renewcommand{\theequation}{S\arabic{equation}}%
\setcounter{section}{0}
\renewcommand{\thesection}{S\arabic{section}}%
\setcounter{page}{1}
% \onecolumngrid
\begin{center}
    \textbf{\large Supplementary Information for 
    ``Demonstrating quantum error mitigation on logical qubits''
    }\\[.2cm]
\end{center}

% \section{Appendix}
\section{Zero-noise extrapolation formulas}
\label{sec:ZNEformulas}

First, we introduce some notations and briefly review NISQ and FTQC circuits. Here, FTQC circuits refer to quantum circuits with mid-circuit state preparation and measurement, feedback and post-selection. Then, we show how to express an FTQC circuit with an example. Finally, we generalize the expression to arbitrary FTQC circuits. With the expression, we derive the formula for ZNE. 

{\bf Notations.} We use $X_i,Y_i,Z_i$ to denote Pauli operators on qubit-$i$, and we use $\openone$ to denote the identity operator. Given an operator $V$, we use $[V]$ to denote a completely positive map $[V]\bullet = V\bullet V^\dag$. 

We use three types of primitive operations to implement quantum computing: gate, state preparation and measurement. We can denote them with completely positive maps. An ideal gate is denoted by $[U]$, where $U$ is a unitary operator. An ideal state preparation is denoted by $\mathcal{A}_{P,i} = [(\openone+P_i)/2] + [P'_i][(\openone-P_i)/2]$, which prepares qubit-$i$ in the eigenstate of $P = X,Y,Z$ with the eigenvalue $+1$. Here, $P'$ is an arbitrary Pauli operator different from $P$, and $(\openone\pm P_i)/2$ is the projection operator onto eigenstates of $P_i$ with the eigenvalue $\pm 1$. An ideal measurement is denoted by $\mathcal{B}_{P,i}(\mu) = [(\openone+\mu P_i)/2]$, which is a measurement on qubit-$i$ in the $P = X,Y,Z$ basis. Here, $\mu = \pm 1$ is the measurement outcome. The maps $[U]$ and $\mathcal{A}_{P,i}$ are trace-preserving; $\mathcal{B}_{P,i}(\mu)$ is not, but $\mathcal{B}_{P,i}(+1)+\mathcal{B}_{P,i,}(-1)$ is trace-preserving. 

When errors occur stochastically in an operation, the actual operation is in the form $\mathcal{M} = (1-p)\mathcal{M}^I + p\mathcal{M}^E$, where $\mathcal{M}^I = [U],\mathcal{A}_{P,i},\mathcal{B}_{P,i}(\mu)$ is the ideal operation, $p$ is the probability of errors, and $\mathcal{M}^E$ is a completely positive map denoting the operation with errors. 

Now, we take the Pauli error model as an example, which is a practical mode of errors in quantum computing. In the Pauli error model, the process generating Pauli errors is a map in the form $\mathcal{N} = (1-p)[\openone] + p\mathcal{E}$, where $p$ is the error probability, and $\mathcal{E}$ is a trace-preserving completely positive map denoting errors. If we neglect crosstalk, errors only happen on the qubit subset that an operation acts on, called the support. For a single-qubit operation on qubit-$i$, the noise map is $\mathcal{N}_1 = (1-p)[\openone] + (p_X[X_i]+p_Y[Y_i]+p_Z[Z_i])$, where $P_X,P_Y,P_Z$ are probabilities of corresponding Pauli errors, and $p = p_X+p_Y+p_Z$ is the total error probability. Similarly, for a two-qubit gate on qubits $i$ and $j$, the noise map is $\mathcal{N}_2 = (1-p)[\openone] + \sum_{P=X,Y,Z}(p_{PI}[P_i]+p_{IP}[P_j]) + \sum_{P,P'=X,Y,Z}p_{PP'}[P_iP'_j]$, where $p = \sum_{P=X,Y,Z}(p_{PI}+p_{IP}) + \sum_{P,P'=X,Y,Z}p_{PP'}$.  To include crosstalk, the general noise map is in the form $\mathcal{N} = (1-p)[\openone] + \sum_P p_P [P]$, where $P$ is a Pauli operator, $p_P$ is the probability of the Pauli error $P$, and $p = \sum_P p_P$. Here, the Pauli operator $P$ is either a single-qubit Pauli operator or an arbitrary product of single-qubit operators, and it may act on qubits out of operation's support. In this general Pauli-error noise map, the erroneous map is $\mathcal{E} = p^{-1}(\sum_P p_P [P])$. 

In the Pauli error model, the actual operation of a gate is in the form $\mathcal{M} = \mathcal{N}\mathcal{M}^I$, where $\mathcal{M}^I = [U]$ is the ideal gate. For a state preparation operation, the actual operation is also in the form $\mathcal{M} = \mathcal{N}\mathcal{M}^I$, where $\mathcal{M}^I = \mathcal{A}_{P,i}$ is the ideal state preparation. For a measurement, the actual operation is in the form $\mathcal{M}(\mu) = \mathcal{M}^I(\mu)\mathcal{N}$ (the noise occur before the ideal operation this time), where $\mathcal{M}^I(\mu) = \mathcal{B}_{P,i}(\mu)$ is the ideal measurement. Substituting the expression of the noise map $\mathcal{N}$, we can find that actual operations are in the form $\mathcal{M} = (1-p)\mathcal{M}^I + p\mathcal{M}^E$, where $\mathcal{M}^E = \mathcal{E}\mathcal{M}^I$ for gates and state preparation, and $\mathcal{M}^E = \mathcal{M}^I \mathcal{E}$ for measurement. 

{\bf NISQ circuits and FTQC circuits.} In an NISQ circuit, qubits are prepared at the beginning and measured at the end of the circuit, and gates are applied between the state preparation and measurement. Such circuits are insufficient for the usual protocols of FTQC. 

In error correction, we usually detect errors using measurements. Typically, we repeatedly measure a set of carefully chosen Pauli operators, stabilizer operators of the error correction code. Then, we send the measurement outcomes to a classical algorithm called a decoder. With the algorithm, we compute a set of Pauli gates. By applying the Pauli gates, we can correct errors on qubits. Therefore, the error correction is a typical feedback process. 

We need post-selection when the magic state is used in quantum computing. Because of the error correction code, we cannot directly apply all necessary gates on logical qubits. The magic state represents a promising protocol to complete the universal gate set. In the magic-state protocol, we prepare some resource states on logical qubits and then improve their fidelities in certain distillation circuits. With the resource states distilled, we can use them to implement gates that cannot be implemented directly. The distillation circuits improve the fidelity through post-selection: the resource state is kept or discarded depending on measurements in the distillation circuit; if discarded, we need to re-prepare the state and try distillation for another round. We remark that the preparation, distillation and utilization of magic states also require feedback operations. 

\begin{figure}[tbp]
\centering
\includegraphics[width=3.5 in]{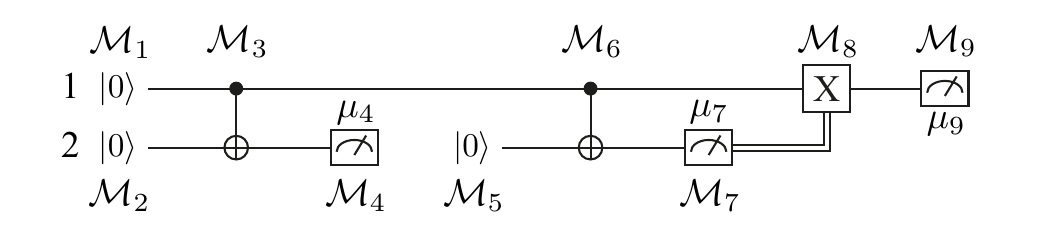}
\caption{
An example circuit with feedback and post-selection operations. The circuit succeeds when $\mu_4 = +1$. 
}
\label{fig:circuit_FTQC}
\end{figure}

{\bf An example FTQC circuit.} To illustrate the expression of FTQC circuits using completely positive maps, we take the circuit in Fig.~\ref{fig:circuit_FTQC} as an example, which includes feedback and post-selection operations. There are three state preparation operations and three measurements in the circuit. Depending on the measurement outcome $\mu_7$, the number of gates is either two or three. 

Using completely positive maps, we can express the ideal circuit in the following form 
\begin{eqnarray}
&& \mathcal{M}^I(\mu_4,\mu_7,\mu_9) \notag \\
&=& \delta_{\mu_4,+1}\mathcal{M}_9^I(\mu_9)\mathcal{M}_8^I(\mu_7)\mathcal{M}_7^I(\mu_7)\mathcal{M}_6^I \notag \\
&&\times \mathcal{M}_5^I\mathcal{M}_4^I(\mu_4)\mathcal{M}_3^I\mathcal{M}_2^I\mathcal{M}_1^I.
\end{eqnarray}
Here, $\mathcal{M}_1^I = \mathcal{A}_{Z,1}$ and $\mathcal{M}_2^I = \mathcal{M}_5^I = \mathcal{A}_{Z,2}$ are state preparation operations, $\mathcal{M}_3^I = \mathcal{M}_6^I = [U]$ are controlled-NOT gates, $\mathcal{M}_4^I(\mu_4) = \mathcal{B}_{Z,2}(\mu_4)$, $\mathcal{M}_7^I(\mu_7) = \mathcal{B}_{Z,2}(\mu_7)$ and $\mathcal{M}_9^I(\mu_9) = \mathcal{B}_{Z,1}(\mu_9)$ are measurements, $\mathcal{M}_8^I(\mu_7) = \delta_{\mu_7,+1}[\openone] + \delta_{\mu_7,-1}[X_1]$ is the measurement-dependent gate in the feedback operation, and $\delta_{\mu_4,+1}$ denotes the post-selection operation. 

With noise, the map of the circuit becomes 
\begin{eqnarray}
&& \mathcal{M}(\mu_4,\mu_7,\mu_9) \notag \\
&=& \delta_{\mu_4,+1}\mathcal{M}_9(\mu_9)\mathcal{M}_8(\mu_7)\mathcal{M}_7(\mu_7)\mathcal{M}_6 \notag \\
&&\times \mathcal{M}_5\mathcal{M}_4(\mu_4)\mathcal{M}_3\mathcal{M}_2\mathcal{M}_1,
\end{eqnarray}
where operations are in the from $\mathcal{M}_j = (1-p_j)\mathcal{M}_j^I + p_j\mathcal{M}_j^E$. Because the eighth operation is measurement-dependent, it has a measurement-dependent error probability $p_8(\mu_7)$ and erroneous operation $\mathcal{M}_8^E(\mu_7)$: $p_8(+1)$ and $\mathcal{M}_8^E(+1)$ [$p_8(-1)$ and $\mathcal{M}_8^E(-1)$] are the error probability and erroneous operation, respectively, of the gate $\openone$ ($X_1$). 

{\bf General FTQC circuits.} For a general FTQC circuit, we can express the ideal circuit consisting of $N$ operations in the form 
\begin{eqnarray}
\mathcal{M}^I(\tilde{\mu}_N) &=& g(\tilde{\mu}_N)\mathcal{M}_N^I(\tilde{\mu}_{N-1},\mu_N)\cdots \notag \\
&&\times\mathcal{M}_j^I(\tilde{\mu}_{j-1},\mu_j)\cdots\mathcal{M}_2^I(\tilde{\mu}_1,\mu_2)\mathcal{M}_1^I(\mu_1).~~
\end{eqnarray}
Here, $\mathcal{M}_j^I$ is the completely positive map denoting the $j$th operations. We use $\mu_j$ to represent the measurement outcome of the $j$th operation: $\mu_j = \pm 1$ ($\mu_j = 0$) if the $j$th operation is (not) a measurement. If the $j$th operation is a measurement, the map $\mathcal{M}_j^I$ is a function of the measurement outcome $\mu_j$; even if the $j$th operation is not a measurement, we still can express it as a function of $\mu_j$ because $\mu_j$ takes only one value anyway. We use $\tilde{\mu}_j$ to represent a tuple of measurement outcomes from the first to $j$th operations, and we can define it formally with the recursion formula $\tilde{\mu}_j = (\tilde{\mu}_{j-1},\mu_j)$ and the initial value $\tilde{\mu}_0 = ()$. The $j$th operation may depend on measurement outcomes of previous operations since the feedback. Therefore, the map $\mathcal{M}_j^I$ is also a function of $\tilde{\mu}_{j-1}$: If the operation is not a feedback operation, the dependence is trivial, i.e. $\mathcal{M}_j^I(\tilde{\mu}_{j-1},\mu_j) = \mathcal{M}_j^I(\mu_j)$, such as operations with $j\neq 8$ in the example circuit; the eighth operation is the only one with a non-trivial dependence on previous measurement outcomes. The function $g$ describes the post-selection, 
\begin{eqnarray}
g(\tilde{\mu}_N) &=& \begin{cases}
1, & \tilde{\mu}_N\in S; \\
0, & \tilde{\mu}_N\notin S,
\end{cases}
\end{eqnarray}
where $S$ denotes the set of measurement outcomes indicating success. 

For the noisy circuit, its map is in the form 
\begin{eqnarray}
\mathcal{M}(\tilde{\mu}_N) &=& g(\tilde{\mu}_N)\mathcal{M}_N(\tilde{\mu}_{N-1},\mu_N)\cdots \notag \\
&&\times\mathcal{M}_j(\tilde{\mu}_{j-1},\mu_j)\cdots\mathcal{M}_2(\tilde{\mu}_1,\mu_2)\mathcal{M}_1(\mu_1),~~
\label{eq:M}
\end{eqnarray}
where operations are in the from 
\begin{eqnarray}
\mathcal{M}_j(\tilde{\mu}_{j-1},\mu_j) &=& [1-p_j(\tilde{\mu}_{j-1})]\mathcal{M}_j^I(\tilde{\mu}_{j-1},\mu_j) \notag \\
&&+ p_j(\tilde{\mu}_{j-1})\mathcal{M}_j^E(\tilde{\mu}_{j-1},\mu_j).
\end{eqnarray}
If the $j$th operation is a feedback operation that depends on previous measurement outcomes, such as the eighth operation in the example circuit, error probability $p_j$ depends on $\tilde{\mu}_{j-1}$; otherwise, $p_j$ is a constant function; the erroneous operation $\mathcal{M}_j^E$ is similar. 

The expression in Eq. (\ref{eq:M}) illustrates the temporal order of operations that an operation can only depend on measurement outcomes in previous operations. This temporal order is unimportant for error mitigation, and if neglect, we have a simplified expression of a noisy circuit: We use $\boldsymbol{\mu} = \tilde{\mu}_N$ to denote all measurement outcomes in the circuit, then 
\begin{eqnarray}
\mathcal{M}(\boldsymbol{\mu}) &=& g(\boldsymbol{\mu})\mathcal{M}_N(\boldsymbol{\mu})\cdots\mathcal{M}_j(\boldsymbol{\mu})\cdots\mathcal{M}_2(\boldsymbol{\mu})\mathcal{M}_1(\boldsymbol{\mu}),~~~
\end{eqnarray}
where operations are in the from 
\begin{eqnarray}
\mathcal{M}_j(\boldsymbol{\mu}) &=& [1-p_j(\boldsymbol{\mu})]\mathcal{M}_j^I(\boldsymbol{\mu}) + p_j(\boldsymbol{\mu})\mathcal{M}_j^E(\boldsymbol{\mu}).
\end{eqnarray}

{\bf Expansion formula of the observable with errors.} With a circuit, we can evaluate observables that are functions of measurement outcomes, $f(\boldsymbol{\mu})$. For example, if we want to evaluate the observable $Z_1$ in the example circuit, we take $f(\boldsymbol{\mu}) = \mu_9$. In the case of error correction, this function takes into account the last-round error correction on the measurement outcomes. 

The ideal expected value of the observable is 
\begin{eqnarray}
\langle O\rangle_{\rm ideal} &=& \sum_{\boldsymbol{\mu}} f(\boldsymbol{\mu}) \Tr\left[\mathcal{M}^I(\boldsymbol{\mu})\rho\right],
\label{eq:OI}
\end{eqnarray}
where $\rho$ is the initial state, and $\Tr\left[\mathcal{M}^I(\boldsymbol{\mu})\rho\right]$ is the probability of the measurement outcome $\boldsymbol{\mu}$ in the ideal circuit. 

Because of the post-selection, the distribution may not be normalized in general. The expected value in the normalized distribution is 
\begin{eqnarray}
\langle O\rangle_{\rm ideal} &=& \frac{\langle O\rangle_{\rm ideal}}{\langle \openone\rangle_{\rm ideal}},
\end{eqnarray}
and we have the expression of $\langle \openone\rangle_{\rm ideal}$ by substituting $f = 1$ into Eq. (\ref{eq:OI}). In the following, we focus on $\langle O\rangle$, and the result can be applied to $\langle \openone\rangle$. 

With a noisy circuit, the expected value of the observable is 
\begin{eqnarray}
\langle O\rangle &=& \sum_{\boldsymbol{\mu}} f(\boldsymbol{\mu}) \Tr\left[\mathcal{M}(\boldsymbol{\mu})\rho\right],
\label{eq:O}
\end{eqnarray}
where $\Tr\left[\mathcal{M}(\boldsymbol{\mu})\rho\right]$ is the probability of the measurement outcome $\boldsymbol{\mu}$ in the noisy circuit. To explicitly express the observable as a function of error probabilities, we rewrite each actual operation in the circuit as 
\begin{eqnarray}
\mathcal{M}_j &=& \mathcal{M}_j^{(0)} + p_j\mathcal{M}_j^{(1)} = \sum_{b_j = 0,1} p_j^{b_j}\mathcal{M}_j^{(b_j)},
\end{eqnarray}
where $\mathcal{M}_j^{(0)} = \mathcal{M}_j^I$ is the ideal operation, and 
\begin{eqnarray}
\mathcal{M}_j^{(1)} &=& \mathcal{M}_j^E - \mathcal{M}_j^I
\end{eqnarray}
is the deviation from the ideal operation. Substituting this expression of $\mathcal{M}_j$, we have 
\begin{eqnarray}
\langle O\rangle &=& \sum_{\boldsymbol{\mu}}  \sum_{b_1,b_2,\ldots,b_N=0,1} \left[\prod_{j=1}^N p_j^{b_j}(\boldsymbol{\mu})\right] \notag \\
&& \times a(b_1,b_2,\ldots,b_N;\boldsymbol{\mu}),
\end{eqnarray}
where 
\begin{eqnarray}
&& a(b_1,b_2,\ldots,b_N;\boldsymbol{\mu}) \notag \\
&=& f(\boldsymbol{\mu}) g(\boldsymbol{\mu}) \Tr\left[\mathcal{M}_N^{(b_N)}(\boldsymbol{\mu})\cdots\mathcal{M}_1^{(b_1)}(\boldsymbol{\mu})\rho\right]. 
\end{eqnarray}

Eventually, we can rewrite the expression as 
\begin{eqnarray}
\langle O\rangle = \langle O\rangle_{\rm ideal} + \sum_{k=1}^N a_k,
\end{eqnarray}
where 
\begin{eqnarray}
a_k &=& \sum_{\boldsymbol{\mu}} \sum_{j_1<j_2<\cdots<j_k} \notag \\
&& p_{j_1}(\boldsymbol{\mu})p_{j_2}(\boldsymbol{\mu})\cdots p_{j_k}(\boldsymbol{\mu}) a(j_1,j_2,\ldots,j_k;\boldsymbol{\mu}),
\end{eqnarray}
and 
\begin{eqnarray}
&& a(j_1,j_2,\ldots,j_k) \notag \\
&=& a(b_1,b_2,\ldots,b_N)\vert_{b_j=1\text{ iff } j\in\{j_1,j_2,\ldots,j_N\}}.
\end{eqnarray}
Here, we have used that $\sum_{\boldsymbol{\mu}} a(0,0,\ldots,0;\boldsymbol{\mu}) = \langle O\rangle_{\rm ideal}$. 

In ZNE, we mitigate errors by boosting the error probability. If we boost errors in each operation with the same factor of $r$ (i.e. $p_j\rightarrow rp_j$ for all $j$), the computing result becomes a polynomial function of $r$, $\langle O\rangle(r) = \langle O\rangle_{\rm ideal} + \sum_{k=1}^N a_k r^k$. This formula holds for both NISQ and FTQC circuits. 

\section{Correlations in logical errors of qLDPC codes}

\begin{figure}[!htbp]
\centering
\includegraphics[width=3.5 in]{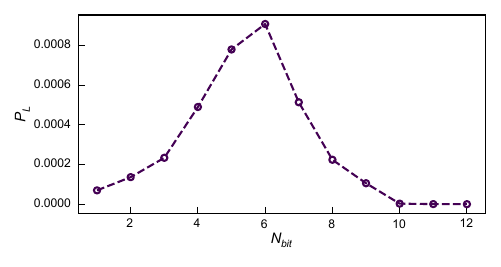}
\caption{
{\bf Numerical results of multi-qubit correlations in a bivariate bicycle code.}  We take the $[[72,12,6]]$ code in the bivariate bicycle family and a physical error rate of $p=0.002$ per operation as an example. Using Monte Carlo simulation, we evaluate the number of logical qubits $N_{bit}$ that are affected by logical errors simultaneously in each trial. The probability of each qubit number is ploted. 
}
\label{fig:LDPC}
\end{figure}

Unlike the surface code, many qLDPC codes encode a large number of logical qubits in a single block of physical qubits. Errors in these logical qubits could be significantly correlated, tending to occur simultaneously on multiple logical qubits, as illustrated in Fig.~\ref{fig:LDPC}. In the $[[72,12,6]]$ code \cite{bravyi_high-threshold_2024}, the probability that $4$ to $7$ logical qubits fail simultaneously is higher than other cases, indicating significant correlations. For such correlated errors, measuring their error rates is extremely challenging. Even assuming Pauli errors, obtaining a model of correlated errors requires measuring $4^k-1$ error rates, where $k$ is the number of logical qubits in a block. 

\section{Device performance}

\begin{figure}[!htbp]
\centering
\includegraphics[width=3.5 in]{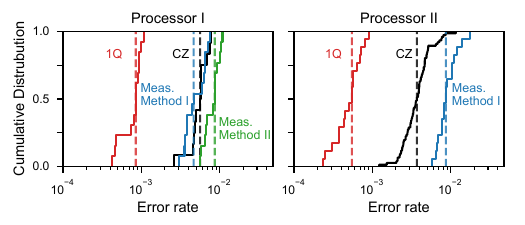}
\caption{
{\bf Cumulative distributions of operation error rates measured on Processor~\MakeUppercase{\romannumeral 1} and Processor~\MakeUppercase{\romannumeral 2}.} Red: Pauli errors for simultaneous single-qubit gates. Black: Pauli errors for simultaneous CZ gates. Blue: average identification errors for measurement where additional microwave pulses are applied to excite each qubit from the state $\ket{1}$ to the state $\ket{2}$ before performing the dispersive readout (Method~\MakeUppercase{\romannumeral 1}). Green: average identification errors for conventional measurement (Method~\MakeUppercase{\romannumeral 2}). Dashed lines: median values.
}
\label{fig:figS_processors_errors_info}
\end{figure}

Two quantum processors, named Processor~\MakeUppercase{\romannumeral 1} and Processor~\MakeUppercase{\romannumeral 2},  are used in this work for experiments of repetition code and surface code, respectively.
Processor~\MakeUppercase{\romannumeral 1} consists of a $6 \times 6$ transmon qubit array, from which 13 qubits are utilized in the experiment.
Processor~\MakeUppercase{\romannumeral 2} has the same architecture as Processor~\MakeUppercase{\romannumeral 1}, but with a qubit array of $11 \times 11$.
In the distance-3 surface code experiments, 17 qubits are used.
The median $T_1$ and $T_{2}^{\rm SE}$ times for the qubits used in Processor~\MakeUppercase{\romannumeral 1} (Processor~\MakeUppercase{\romannumeral 2}) are $124$~$\mu$s and $11$~$\mu$s ($128$~$\mu$s and $16$~$\mu$s), respectively. 
In this work, we adopt two different readout schemes: (1) conventional measurement for multi-round experiments; (2) using state $\ket{2}$ to reduce measurement errors for one-round experiments.
The gate and measurement errors of two processors are visualized in Fig.~\ref{fig:figS_processors_errors_info}.

\begin{table*}[!htbp]
    \setlength{\tabcolsep}{3pt}
    \renewcommand{\arraystretch}{1.5}
    \centering
    \caption{\label{tab:device_parameters} {\bf Qubit parameters, coherence properties and gate performance for Processor~\MakeUppercase{\romannumeral 1} and Processor~\MakeUppercase{\romannumeral 2}.}}
    \begin{tabular}{c|ccc|ccc}
        \hline \hline
         & & Processor~\MakeUppercase{\romannumeral 1} & & & Processor~\MakeUppercase{\romannumeral 2} & \\
        % \hline
        Parameter & Median & Mean & Stdev. & Median & Mean & Stdev. \\
        \hline
        Qubit idle frequency, $\omega_{\rm idle}/2\pi$ (GHz) & 4.000 & 3.962 & 0.128 & 4.123 & 4.120 & 0.131 \\
        Qubit anharmonicity, $\alpha/2\pi$ (MHz) & -197.3 & -197.1 & 1.6 & -214.6 & -214.7 & 2.3 \\
        Readout frequency, $\omega_{\rm r}/2\pi$ (GHz) & 6.316 & 6.299 & 0.095 & 6.416 & 6.416 & 0.086 \\
        % Resonator linewidth, $\kappa/2\pi$ (MHz) & \\
        Energy relaxation time, $T_1$ ($\mu$s) & 124.36 & 111.82 & 33.22 & 127.89 & 132.44 & 24.32 \\
        Spin-echo dephasing time, $T_{2}^{\rm SE}$ ($\mu$s) & 11.34 & 11.72 & 2.35 & 15.55 & 18.18 & 7.81 \\
        Readout error $e_{\rm r}$ using Method~\MakeUppercase{\romannumeral 1} / Method~\MakeUppercase{\romannumeral 2} (\%)& 0.47 / 0.87 & 0.52 / 0.82 & 0.16 / 0.18 & 0.87 / - & 0.94 / - & 0.30 / - \\
        1Q XEB Pauli error, $e_1$ (\%)& 0.085 & 0.079 & 0.021 & 0.055 & 0.052 & 0.019 \\
        2Q XEB Pauli error, $e_2$ (\%)& 0.56 & 0.56 & 0.14 & 0.37 & 0.40 & 0.17 \\
        \hline \hline
    \end{tabular}
\end{table*}

\section{Generating circuit instances and calculating the expectations}\label{sec:generating_circuits}

To controllably amplify existing errors, we choose to insert an operation drawn from the list of $\{I, X, Y, Z\} $ during the operational stage for each qubit. 
The most intuitive way to generate adequate circuit instances for the estimation of $\langle Z_{\rm L} \rangle$ (or $\langle X_{\rm L} \rangle$) --- randomly selecting the operation with assigned probabilities for each circuit instance --- can be inefficient.
For example, in the scenario of using a distance-3 (7) repetition code, around $80\%$ ($60\%$) of circuit instances are devoid of any error injection when $p=3.6\%$ and $r=1$.
In addition, circuit instances with the presence of multiple injected errors, which can potentially lead to the failure of quantum error correction and contribute to the infidelity of estimation, are unlikely to be chosen.
Alternatively, we can iterate over all possible circuit instances and numerically construct the desired circuit list by calculating the weighted average. 
For example, in the experiment related to Fig.~2 of the main text, all $4^3$ circuit instances were experimentally implemented for each value of $r$, and the resulting data were processed to estimate the expectations of $Z_0$. However, in repetition and surface codes, the large number of circuit instances makes full implementation impractical. To address this challenge, we employ a method to select the most likely circuit instances as follows.

First, we determine the total number of circuit instances, which scales proportionally with the code distance to ensure more accurate estimations (see Table~\ref{tab:circuit_number}). Second, we consider the case of injecting $k$ errors into the circuit and calculate the required number of circuit instances based on the error-injection probability $rp$. Circuit instances are then randomly selected without replacement until the desired number is reached or all possible instances have been selected. The number of remaining circuit instances to be determined is updated accordingly. Finally, the second and third steps are iteratively repeated, starting with $k=0$ and continuing until the total number is reached.

The above procedure assumes perfect operations and measurements. In practice, however, when the number of possible circuit instances for injecting $k$ errors is small --- such as in the case of $k=0$ --- the statistical error due to finite measurement shots can proportionately impact the overall estimation accuracy, scaling with the assigned weight. To reduce this effect, we introduce a key modification to the procedure: the circuit number for each $k$ is required to exceed a predefined threshold, such as $1\%$ of the total circuit number. Equivalently, this modification increases the number of measurement shots, thereby reducing statistical errors and improving the accuracy of the estimation.

\begin{table*}[!htbp]
    \setlength{\tabcolsep}{12pt}
    \renewcommand{\arraystretch}{1.5}
    \centering
    \caption{\label{tab:circuit_number} {\bf Number of circuit instances used to estimate the expectation values.}}
    \begin{tabular}{c|rrrr}
        \hline \hline
         & 1-round & 2-round & 3-round & 4-round \\
        \hline
        distance-3 repetition code & $1000$ & $1000$ & $1500$ & $2000$ \\
        distance-5 repetition code & $3500$ & $5000$ & $5000$ & $6000$ \\
        distance-7 repetition code & $6000$ & $6000$ & $6000$ & $7000$ \\
        \hline
        distance-3 surface code    & $4000$    & - & - & - \\
        \hline
        \hline
    \end{tabular}
\end{table*}

We estimate the expectation values, such as $\langle Z_{\rm L} \rangle$ and $\langle X_{\rm L} \rangle$, by averaging measured results across all generated circuit instances with the respective weights.
Specifically, we denote $O_{k,c,s}$ as the corresponding outcome with $k$ errors are injected into the circuit, with $c$ representing the circuit index and $s$ the index of measurement shot. 
The expectation value is then calculated as
\begin{equation}
    \overline{O} = \frac{1}{S}\sum_{k, c, s} P(k) \frac{O_{k, c, s}}{C(k)},
    \label{eq:average_mean}
\end{equation}
where $S$ represents the number of measurement shots for each circuit instance, $C(k)$ denotes the number of chosen circuits with $k$ errors injected, and $P(k)$ is the respective weight.
The value of $S$ for experiments of repetition code and surface code is $150$.

\section{More experimental details} % \sout{results}

\begin{figure}[!htbp]
\centering
\includegraphics[width=3.5 in]{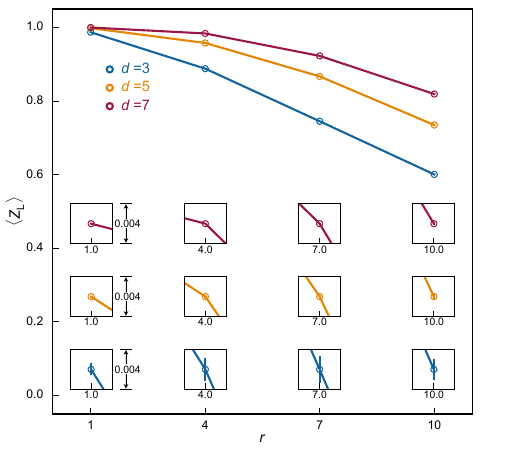}
\caption{
{\bf Standard error of the measured $\langle Z_{\rm L} \rangle$.}
All insets have a spread of $0.004$ along the $y$-axis, with the data points positioned at the center of the figures. 
}
\label{fig:figS_standard_errors}
\end{figure}

{\bf Standard error of the measured expectations.}
As mentioned in Section~\ref{sec:generating_circuits}, $S$ measurement shots are taken for each circuit. 
It allows us to directly calculate the standard deviation of $\overline{O}$ (see Eq.~\ref{eq:average_mean}) without requiring additional experimental effort.
First, we calculate the average expectation value for each shot index $s$ as
\begin{equation}
    \overline{O}_{s} = \sum_{k, c} P(k) \frac{O_{k, c, s}}{C(k)},
\end{equation}
where the sum is taken over the error count $k$ and circuit instance $c$.
Next, the unbiased standard deviation of $\overline{O}$ can be obtained by
\begin{equation}
    \sigma\left[\overline{O}\right] = \sqrt{\frac{1}{S(S-1)} \sum_{s} \left(\overline{O}_{s} - \overline{O}\right)^2}.
\end{equation}
We note that the standard errors are relatively small for both the repetition code and the surface code, making them difficult to visualize clearly in the main figures. 
See Fig.~\ref{fig:figS_standard_errors} for an enlarged view of the standard errors for the one-round repetition code.

\begin{figure}[!htbp]
\centering
\includegraphics[width=3.5 in]{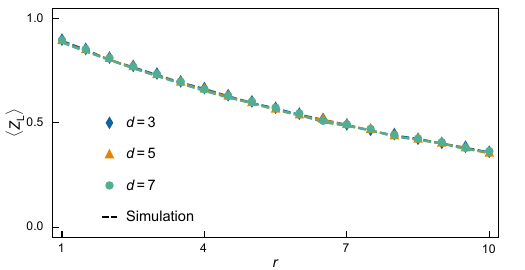}
\caption{
{\bf Uncorrected results of the one-round repetition code.}
For different repetition code distances, the uncorrected expectation values of $Z_{\rm L}$ remain nearly the same.
Dashed lines: noisy numerical simulations.
}
\label{fig:figS_uncorrected_expectation_values}
\end{figure}

{\bf Uncorrected expectation values for the repetition code.}
In Fig.~\ref{fig:figS_uncorrected_expectation_values}, we observe that the uncorrected expectation values for different repetition code distances nearly overlap.
The slight differences observed for large noise scaling factors are mainly due to the randomness of the circuit instances, which also influence the simulation results.

\begin{figure}[!htbp]
\centering
\includegraphics[width=3.5 in]{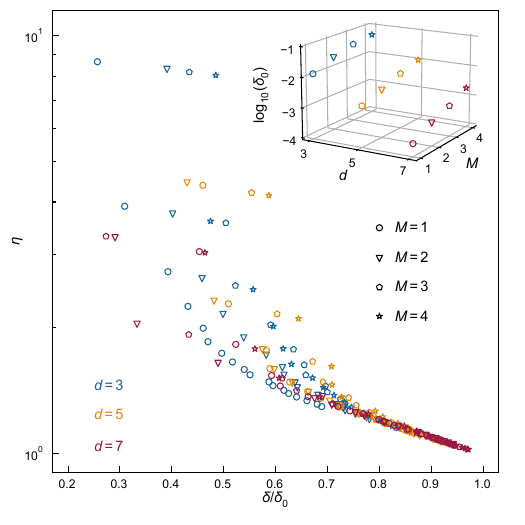}
\caption{
{\bf Relative bias and sampling overhead of the ZNE.}
For the implementation of ZNE, we use $K=1$. Note that $\delta/\delta_0$ is presented in a linear scale, whereas $\eta$ is shown in a logarithmic scale. 
Inset: biases without ZNE.
}
\label{fig:figS_relative_bias}
\end{figure}

{\bf Investigating the scalability of ZNE with a fixed error rate per parity-check round.} 
Unlike in the main text, where the unit error probability for each $M$ is adjusted to maintain a fixed total error rate (see Table~\ref{tab:unit_error_probability}), in this study, the unit error probability in each parity-check round is fixed at $p=3.6\%$ and then the performance of ZNE is evaluated. As depicted in Fig.~\ref{fig:figS_relative_bias}, the relative bias, represented by $\delta/\delta_0$, consistently remains below 1, confirming the effectiveness of ZNE. The ZNE results demonstrate consistent trends for different code distances and parity-check round numbers, with a larger code distance resulting in a smaller sampling overhead. Additionally, the sampling overhead remains nearly unchanged for different numbers of parity-check rounds. These findings demonstrate the scalability of ZNE for error correction circuits, even with increasing circuit complexity.

\begin{table}[!htbp]
    \setlength{\tabcolsep}{6pt}
    \renewcommand{\arraystretch}{1.5}
    \centering
    \caption{{\bf Unit error probabilities.} For each $M$, the unit error probability is chosen so that the total error rate is fixed.}
    \begin{tabular}{c|cccc}
        \hline \hline
        Round, $M$ & 1 & 2 & 3 & 4 \\
        \hline
        Unit error probability, $p$ ($\%$) & 13.6 & 9.4 & 7.2 & 5.7 \\
        \hline \hline
    \end{tabular}
    \label{tab:unit_error_probability}
\end{table}

\begin{figure*}[!htbp]
\centering
\includegraphics{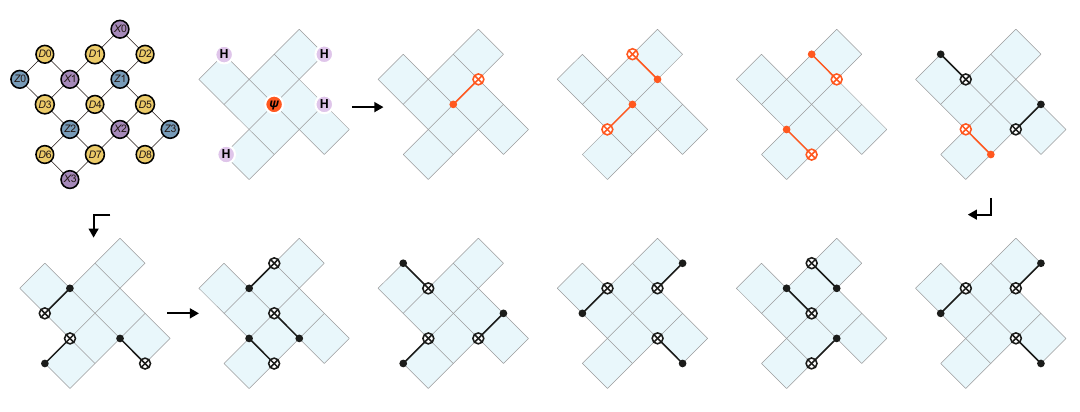}
\caption{
{\bf Circuit for the initialization of an arbitrary logical state.}
The preparation of an arbitrary logical state $\ket{\psi_{\rm L}} = \alpha \ket{0_{\rm L}} + \beta \ket{1_{\rm L}}$ proceeds as follows. First, a single-qubit gate initializes data qubit $D4$ in the state $|\psi\rangle = \alpha|0\rangle + \beta|1\rangle$, with Hadamard gates applied to four representative qubits. Next, six CNOT gates (highlighted in red) are used to create a GHZ-like state --- $\alpha|000\rangle + \beta|111\rangle$ --- on the vertically aligned data qubits $D1$, $D4$, and $D7$. Finally, multiple layers of CNOT gates are sequentially applied to construct the desired logical state. In particular, for the preparation of state $\ket{0_{\rm L}}$, the six CNOT gates highlighted in red are unnecessary and can be omitted.
}
\label{fig:figS_initialization_circuit_for_surface_code}
\end{figure*}

{\bf Preparation of an arbitrary logical state in the distance-3 surface code.} 
In this work, an arbitrary logical state in the distance-3 surface code is prepared using the circuit illustrated in Fig.~\ref{fig:figS_initialization_circuit_for_surface_code}. For specific cases, such as the initialization of states $ \ket{0_{\rm L}} $ and $ \ket{+_{\rm L}} $, the circuit can be further simplified, as described in Ref.~\cite{ZhaoPan2022}. Additionally, the implementation of state initialization in a configuration where two-qubit gates can be performed between nearest-neighbor data qubits is detailed in Ref.~\cite{SatzingerRoushan2021}.

\begin{figure*}[!htbp]
\centering
\includegraphics{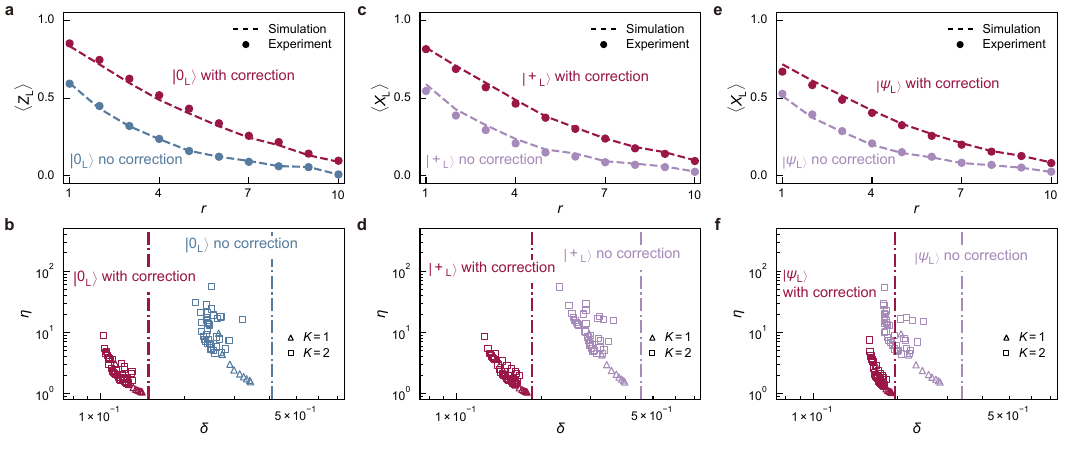}
\caption{
{\bf Experimental results of ZNE for the $\ket{0_{\rm L}}$ and $\ket{+_{\rm L}}$ states in the distance-3 surface code.}
{\bf a}--{\bf b}, Measured expectation values of the $Z_{\rm L}$ observable for the logical state $\ket{0_{\rm L}}$ and corresponding scatter plots of the bias $\delta$ and sampling overhead $\eta$. 
{\bf c}--{\bf f}, Similar to panels {\bf a} and {\bf b}, but with the observable being $X_{\rm L}$ and the initial states being $\ket{+_{\rm L}}$ (panels {\bf c} and {\bf d}) and $|\psi_{\rm L}\rangle=\cos\frac{\pi}{6}|0_{\rm L}\rangle + \sin\frac{\pi}{6}|1_{\rm L}\rangle$ (panels {\bf e} and {\bf f}).
}
\label{fig:figS_zero_and_plus_state_for_surface_code}
\end{figure*}

{\bf More ZNE results in the distance-3 surface code. }
The expectation values of the $Z_{\rm L}$ and $X_{\rm L}$ observables are measured as a function of the noise scaling factor $r$ for the states $\ket{0_{\rm L}}$ and $\ket{+_{\rm L}}$, respectively (Fig.~\ref{fig:figS_zero_and_plus_state_for_surface_code}). 
These results are consistent with those obtained for $\ket{\psi_{\rm L}}$ in the main text but specifically address either bit-flip error or phase-flip error. The measured values of the $X_{\rm L}$ observable, with the initial state set to $|\psi_{\rm L}\rangle=\cos\frac{\pi}{6}|0_{\rm L}\rangle + \sin\frac{\pi}{6}|1_{\rm L}\rangle$, are also shown in the figure.
All $\delta$-$\eta$ scatter plots demonstrate the effectiveness of ZNE in the surface code.

{\bf Numerical simulations as benchmark.}
We conduct numerical simulations using two popular open-source frameworks: Stim~\cite{Gidney2021} and Qiskit~\cite{qiskit2024}. 
For simulations of the repetition code, where the initial preparation of the logical state ($\ket{0_{\rm L}}=\ket{0\cdots0}$) is nearly perfect, errors introduced by the gate and measurement imperfections are then modeled using a depolarizing channel and a bit-flip channel, respectively. The error rates for these models are experimentally calibrated (see Fig.~\ref{fig:figS_processors_errors_info} and Table~\ref{tab:device_parameters}). 
In contrast, for simulations of the surface code, where initial states are imperfectly generated via multiple layers of single-qubit and two-qubit gates, the introduced imperfection is simplified by a single-qubit depolarizing channel uniformly applied to each qubit. The depolarizing rate is calibrated by matching the expectation value of an observable with error injection disabled in the circuit. In this work, $\ket{0_{\rm L}}$ is used as the initial state, and $Z_{\rm L}$ as the observable, resulting in a calibrated depolarizing rate of $0.075$. As shown in Fig.~4 of the main text, this noise model closely aligns with the experimental data.

\section{Protocol}

To implement the $K$th-order ZNE, we use $K+1$ data points from experiments, $\{(r_0,\langle O\rangle(r_0)),(r_1,\langle O\rangle(r_1)),\ldots, (r_K,\langle O\rangle(r_K))\}$. Here, $\langle O\rangle(r)$ denotes the expected value of the observable when error probabilities are amplified by a factor of $r$, and $r_0 = 1$ corresponds to the case without noise amplification. Using these data points and the fitting formula $\langle O\rangle(r) = \langle O\rangle_{\rm em} + \sum_{k=\lceil d/2\rceil}^{\lceil d/2\rceil+K-1} a_k' r^k$, we can compute $\langle O\rangle^I$ as 
\begin{eqnarray}
\langle O\rangle_{\rm em} = \sum_{k=0}^{K} b_k \langle O\rangle(r_k)
\label{eq:Oem}
\end{eqnarray}
where $b_k = \mathrm{adj}(V)_{0k}$, and $V$ is a Vandermonde-like matrix with elements 
\begin{eqnarray}
V_{ij} = \begin{cases}
1, & j = 0, \\
r_i^{\lceil d/2\rceil+j-1}, & j > 0.
\end{cases}
\end{eqnarray}

We characterize the performance of ZNE using two metrics: the bias $\delta = \vert \langle O\rangle_{\rm em} - \langle O\rangle_{\rm ideal}\vert$ characterizes the accuracy, and the sampling overhead $\eta$ characterizes the cost. Next, we focus on the overhead $\eta$. Suppose $N_{tot}$ is the total number of circuit shots used to acquire the $K+1$ data points. To reduce the variance in ZNE, we allocate the shots among the data points according to the importance sampling principle. Specifically, the number of shots used to measure $\langle O\rangle(r_k)$ is $N_k = \frac{\vert b_k\vert }{\sum_{k=0}^{K}\vert b_k\vert }N_{tot}$. The variance in the measurement of $\langle O\rangle(r_k)$ is then given by $\sigma^2_k = \frac{1-\langle O\rangle^2(r_k)}{N_k}$, and the variance of $\langle O\rangle_{\rm em}$ is thus
$\sigma^2_{\rm em} = \sum_{k=0}^{K}\vert b_k\vert ^2\sigma^2_k$. Here we have assumed that the observable $O$ is a Pauli operator. Without ZNE, if the $N_{tot}$ shots are used to measure $\langle O\rangle(r_0)$, the variance is $\sigma^2_{\rm raw}=\frac{1-\langle O\rangle^2(r_0)}{N_{tot}}$. The sampling overhead reads $\eta = \frac{\sigma^2_{\rm em}}{\sigma^2_{\rm raw}}$, which is the ratio of the two variances. 

\section{Application to large-scale logical circuits}

We take the surface code as an example to illustrate that our method can be applied to logical circuits at large scale. Assume we have a logic target circuit consisting of $N$ logical gates. Similar to physical operations (see Sec. \ref{sec:ZNEformulas}), the $j$th logical gate reads 
\begin{eqnarray}
\mathcal{M}_j = (1-P_j)\mathcal{M}_j^I+P_j\mathcal{M}_j^E,
\end{eqnarray}
where $P_j$ represents the logical error rate of the gate. We remark that the logical error rate depends on the physical error rate and, therefore, is a function of the noise amplification factor $r$. 

The expected value of the observable $O$ computed with the circuit is 
\begin{eqnarray}
\langle O\rangle &=& \Tr (O\mathcal{M}_N\cdots\mathcal{M}_2\mathcal{M}_1\rho) \notag \\
&=& (1-\sum_{j=1}^NP_j)\langle O\rangle_{\rm ideal} + \sum_{j=1}^NP_j\langle O\rangle_j + R_1,
\end{eqnarray}
where 
\begin{eqnarray}
\langle O\rangle_{\rm ideal} &=& \Tr (O\mathcal{M}_N^I\cdots\mathcal{M}_2^I\mathcal{M}_1^I\rho)
\end{eqnarray}
is the ideal result, 
\begin{eqnarray}
\langle O\rangle_j &=& \Tr (O\mathcal{M}_N^I\cdots\mathcal{M}_j^E\cdots\mathcal{M}_2^I\mathcal{M}_1^I\rho)
\end{eqnarray}
is the result when the $j$th logical gate has errors, and the remainder term has the upper bound 
\begin{eqnarray}
R_1 &\leq& \Vert O\Vert\left[\prod_{j=1}^N (1+2P_j) - (1+2\sum_{j=1}^NP_j)\right] \notag \\
&\leq& \Vert O\Vert\left[e^{2P_{tot}} - (1+2P_{tot})\right],
\end{eqnarray}
where $P_{tot} = \sum_{j=1}^NP_j$ is the total error rate. 

Now, we approximate each logical error rate with a polynomial, i.e. 
\begin{eqnarray}
P_j(r) = \sum_{k=\lceil d/2\rceil}^{\lceil d/2\rceil+K-1} a_{j,k} r^k + \Delta_j(r),
\end{eqnarray}
and $\Delta_j(r)$ denotes the error in the approximation. With this approximation, we can re-express the expected value as  
\begin{eqnarray}
\langle O\rangle(r) &=& \langle O\rangle_{\rm ideal} + \sum_{k=\lceil d/2\rceil}^{\lceil d/2\rceil+K-1} a_k r^k \notag \\
&&+ R_1(r) + R_2(r),
\end{eqnarray}
where 
\begin{eqnarray}
a_k = \sum_{j=1}^N (\langle O\rangle_j-\langle O\rangle_{\rm ideal}) a_{j,k}
\end{eqnarray}
and 
\begin{eqnarray}
R_2(r) &=& \sum_{j=1}^N (\langle O\rangle_j-\langle O\rangle_{\rm ideal}) \Delta_j(r).
\end{eqnarray}

Substituting the expression of $\langle O\rangle(r)$ into the $K$th-order extrapolation formula Eq. (\ref{eq:Oem}), we obtain the bias 
\begin{eqnarray}
\delta &=& \left\vert \sum_{k=0}^K b_k[R_1(r_k) + R_2(r_k)] \right\vert \leq \tilde{\delta}_1+\tilde{\delta}_2,
\label{eq:bias}
\end{eqnarray}
where 
\begin{eqnarray}
\tilde{\delta}_1 &=& \Vert O\Vert \sum_{k=0}^K \vert b_k\vert \left[e^{2P_{tot}(r_k)} - \left(1+2P_{tot}(r_k)\right)\right]
\end{eqnarray}
and 
\begin{eqnarray}
\tilde{\delta}_2 &=& 2N\Vert O\Vert \left\vert \sum_{k=0}^K b_k \Delta_j(r_k) \right\vert
\end{eqnarray}
are upper bounds of contributions by $R_1$ and $R_2$ terms, respectively. The first term $\tilde{\delta}_1$ is the error due to the second-order contribution of logical error rates, i.e. 
\begin{eqnarray}
\tilde{\delta}_1 &\simeq& 2\Vert O\Vert \sum_{k=0}^K \vert b_k\vert P_{tot}(r_k)^2.
\end{eqnarray}
If we choose a sufficiently large code distance such that $P_{tot}(r_k)$ is much smaller than one, the first term is much smaller than the raw bias before error mitigation, which has the upper bound 
\begin{eqnarray}
\tilde{\delta}_0 &=& \Vert O\Vert \left[e^{2P_{tot}(1)} - 1\right] \simeq 2\Vert O\Vert P_{tot}(1).
\end{eqnarray}
The second term $\tilde{\delta}_2$ is the error due to approximating logical error rates $P_j(r)$ with polynomials. In the limit that the physical error rate $p$ approaches zero, the logical error rate behaves as $P_j(r)\propto p^{\lceil d/2\rceil}$ \cite{PhysRevA.87.040301,watson_logical_2014}, suggesting that the $K = 1$ extrapolation is sufficient in the low physical error rate regime. When $p$ is finite, the logical error rate deviates from $P_j(r)\propto p^{\lceil d/2\rceil}$, and we have to adapt the extrapolation function accordingly. %It is important to note that the approximation is more accurate if we choose a larger $K$ in the polynomial extrapolation, however, resulting in a larger sampling overhead $\eta$. Thus, there is a trade-off between the bias and cost, and an appropriate choice of $K$ must be made. 
Next, we use numerical calculations to estimate the bias in the polynomial extrapolation. 

\begin{figure}[!htbp]
\centering
\includegraphics[width=3.5 in]{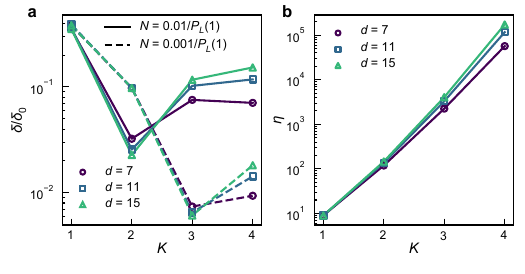}
\caption{
{\bf Numerical results of ZNE on a circuit of $N$ surface-code idle gates.} We take the physical error rate $p = 10^{-3}$, which is amplified to $r_kp$ in ZNE; we choose the amplification factors $r_k = k^{1/\lceil d/2\rceil}$, where $k = 1,2,\ldots,K+1$. The bias before ZNE is $\delta_0 = \vert\langle O\rangle(1)-\langle O\rangle(0)\vert$, where $\langle O\rangle(0)$ is the ideal value. 
}
\label{fig:performance}
\end{figure}

To estimate the bias and cost numerically, we consider a quantum memory circuit consisting of applying $N$ idle gates on a logical qubit and take the logical error rate per gate reported in Ref. \cite{BravyiSimulation2013}. We choose the observable as $Y$, such that the expectation value is affected by both logical $X$ and $Z$ errors. Let $P_L(r)$ be the sum of logical $X$ and $Z$ error rates. The observable expectation value is taken as $\langle O\rangle(r) = [1-2P_L(r)]^N$. Then, we apply ZNE to $\langle O\rangle(r)$, and the results are shown in Fig.~\ref{fig:performance}. 

Although the above numerical results are obtained from quantum memory circuits, we conjecture that general computing circuits yield similar results. Memory circuits are composed of repeatedly applying parity-check measurements on a fixed lattice. Practical protocols of quantum computing in surface codes include, for instance, braiding transformation~\cite{FowlerSurface2012} and lattice surgery~\cite{Horsman_2012}, in which circuits are also composed of repeatedly applying parity-check measurements, however, on a lattice that deforms between certain parity-check cycles. Because the dominant operations are the same in memory circuits and such computing circuits, they should yield similar behavior of logical error rates (as functions of physical error rates and the code distance). In addition to braiding transformation and lattice surgery, universal quantum computing in surface codes also requires magic state injection and distillation~\cite{PhysRevA.71.022316,raussendorf_topological_2007,FowlerSurface2012}. Magic state errors decrease with the level of distillation. As long as the distillation level is adequately high, the post-distillation magic state errors are dominated by logical errors in lattice surgery operations, and the contribution of raw magic state errors (errors in injected magic states before distillation) is negligible; otherwise, we may need to modify the extrapolation function to include the impact of raw magic state errors: suppose the distilled magic state error rate is $\propto p^{L}$, where $L$ is an integer depending on the distillation level, we can add the term $p^L$ to the extrapolation function. 

So far, we have only considered polynomial extrapolation functions. The use of other extrapolation functions could improve the performance. In ZNE on NISQ circuit, it has been shown experimentally that the exponential extrapolation outperforms polynomial extrapolation \cite{endopractical2018,KimKandala2023}. For example, the logical error rate function reported in Ref. \cite{BravyiSimulation2013} is a candidate worth exploring. Given such candidate functions, we can benchmark and verify them through Clifford circuits \cite{Czarnik2021errormitigation,PRXQuantum.2.040330}: we can simulate the Clifford circuits on a classical computer to obtain the ideal result $\langle O\rangle_{\rm ideal}$, such that we can evaluate the bias as well as the cost. In this way, we can compare the candidate the functions and identify the suitable ones.

\end{document}